\documentstyle[aps,prl,twocolumn,overcite]{revtex}
\input epsf
\tighten
\begin{document}
\twocolumn[\hsize\textwidth\columnwidth\hsize\csname @twocolumnfalse\endcsname

\draft
 
%%%%%%%%%%%%%%%%%%%%%%%%%%%%%%%%%%%%%%%%%%%%%%%%%%%%%%%%%%%%%%%%%%%%%%%%%%

\title{Cosmological Imprint of an Energy Component with General Equation-of-State}

\author{ 
R. R. Caldwell,
%\thanks{ Electronic address: caldwell@dept.physics.upenn.edu}, 
Rahul Dave,
%\thanks{ Electronic address: dave@dept.physics.upenn.edu}, 
and 
Paul J. Steinhardt
%\thanks{ Electronic address: steinh@steinhardt.hep.upenn.edu}
}

\address{Department of Physics and Astronomy\\
University of Pennsylvania\\
Philadelphia, PA 19104}

\maketitle

%%%%%%%%%%%%%%%%%%%%%%%%%%%%%%%%%%%%%%%%%%%%%%%%%%%%%%%%%%%%%%%%%%%%%%%%%%
\begin{abstract}

We  examine the possibility that  a significant component of the energy
density of the universe has an equation-of-state different from that of
matter, radiation or cosmological constant ($\Lambda$). An example is a cosmic
scalar field evolving in a potential, but our treatment is more general. 
Including this component alters cosmic  evolution in a way that fits current
observations well. Unlike $\Lambda$, it evolves dynamically and develops
fluctuations, leaving  a distinctive imprint on the  microwave background  
anisotropy and mass power spectrum.
  
\end{abstract} 

\pacs{PACS number(s): 98.80.-k,95.35.+d,98.70.Vc,98.65.Dx,98.80.Cq}
]

Inflationary cosmology  predicts  that the universe is  spatially flat and
that the total energy density of the universe is equal to the critical
density.   This prediction  is consistent with current measurements of the
cosmic microwave background (CMB) anisotropy and  may be verified with high
precision  in the next generation of CMB satellite  experiments.  At the same
time, there is growing observational evidence that the total matter density of
the universe is significantly less than the critical density.\cite{Ostriker}  
If this latter result holds and the CMB anisotropy  establishes that  the
universe is flat, then there must be another contribution to the energy
density of the universe. One candidate that is often considered is a
cosmological constant, $\Lambda$, or vacuum energy density.  The vacuum
density is a spatially uniform, time-independent component.  Cold dark matter
models with a substantial cosmological constant ($\Lambda$CDM)  are among the 
models which best fit existing observational data.\cite{Ostriker} However, it
should be emphasized that the fit depends primarily on the fact that the
models have low matter density and are spatially flat; the fit is not a
sensitive test of whether the additional  energy contribution is vacuum
energy.

In this paper, we consider replacing $\Lambda$ with a dynamical,
time-dependent and spatially inhomogeneous component whose equation-of-state
is different from baryons, neutrinos, dark matter, or radiation. The
equation-of-state of the new component, denoted as $w$, is the ratio of its
pressure to its energy density. This fifth contribution to the cosmic energy 
density, referred to here as ``quintessence" or $Q$-component, is broadly
defined, allowing a spectrum of possibilities including an equation-of-state
which is constant, uniformly evolving or oscillatory.   Examples of a
$Q$-component are fundamental fields (scalar, vector, or tensor) or
macroscopic objects, such as a network of light, tangled cosmic
strings.\cite{Spergel} The analysis in the present paper applies to any
component whose hydrodynamic properties can be mimicked by a scalar field
evolving in a potential which couples to matter only through gravitation.    
In particular, we focus on equations-of-state with $-1<w<0$  because this
range fits current cosmological observations 
best.\cite{Steinhardt,TurnerWhite,CobleEtAl,Hill,BestFitModelsPaper}   This
has motivated several
investigations\cite{Steinhardt,TurnerWhite,SilveraWaga,Wetterich} of
components with $w<0$ in which  a spatially uniform distribution has been
assumed, e.g., a decaying $\Lambda$ or  smooth component.  

In this Letter, we begin by arguing that a smoothly distributed, time-varying
component is unphysical --- it violates the equivalence principle. Hence,
predictions of CMB and mass power spectra  which have not included
fluctuations in the new component are not valid. We outline the general
conditions needed to have $w<0$ consistent with the equivalence principle and
stable against catastrophic gravitational collapse. We show that an evolving
scalar field automatically satisfies these conditions. We then compute the CMB
and mass power spectra for a wide, representative class of models (see Figure
1 and 2a),  taking careful note of the effects of the fluctuations in the
$Q$-component. We show that the fluctuations leave a distinctive signature
that enables a $Q$-component to be distinguished from dark matter and
cosmological constant and makes it possible to resolve its equation-of-state. 
Comparing to  observations of the CMB and large-scale structure (Figure 2b), 
we identify a new spectrum of plausible models that will be targets for future
experiments.

Introducing a dynamical energy component is at least as well motivated by
fundamental physics as introducing a cosmological constant. In fact, the
theoretical prejudice based on fundamental physics is that $\Lambda$ is
precisely zero; if it is non-zero, there is no conceivable mechanism to
explain why the vacuum density should be comparable to the present matter
density, other than arguments based on the anthropic principle. On the other
hand, dynamical fields abound in quantum gravity, supergravity and superstring
models ({\it e.g.}, hidden sector fields, moduli, pseudo-Nambu-Goldstone
bosons), and it may even be possible to utilize the interaction of these
fields with matter to find a natural explanation why the $Q$-component  and
matter have comparable energy densities today.

As noted above, a number of studies\cite{Steinhardt,TurnerWhite,SilveraWaga}
have assumed a ``smooth" (spatially uniform), time-dependent component with
arbitrary equation-of-state (sometimes called $x$CDM) which does not respond 
to the inhomogeneities in the dark matter and baryon-photon-neutrino fluid. In
computing the CMB anistropy, one finds a near-degeneracy\cite{Steinhardt} with
$\Lambda$CDM for a wide range of $w$. In this Letter, we show that the
degeneracy is significantly broken when fluctuations in the $Q$-component are
included. However, it is important to realize that the smooth and
$Q$-scenarios are not competing models.  A smooth, time-evolving component is
ill-defined, since the smoothness  is gauge-dependent,  and unphysical,
because it violates the equivalence principle to ignore the response of the
new component to the inhomogeneities in the surrounding cosmological fluid.
Hence, a fluctuating, inhomogeneous component is the only valid way of
introducing an additional energy component.

There have been some discussions in the literature of an energy component
consisting of a dynamical, fluctuating cosmic scalar field evolving in a
potential,\cite{CobleEtAl,RatraPeebles,FerreiraJoyce} or an energy component
evolving  according to a specific
equation-of-state.\cite{TurnerWhite,SilveraWaga,Chiba}    In the case of a
scalar field, the CMB anisotropy and power spectrum were computed by
Coble {\it et al} for  a cosine potential,\cite{CobleEtAl} and by Ferreira
and Joyce for an exponential potential.\cite{FerreiraJoyce} Here we go beyond
these isolated examples to explore the range of possibilities and the range of
imprints on the CMB and large-scale structure.

The class of cosmological models we consider in this work are spatially flat,
Friedmann-Robertson-Walker (FRW) space-times which contain baryons, neutrinos,
radiation, cold dark matter and the $Q$-component (QCDM models). The
space-time metric is given by $ds^2 = a^2(\eta) ( -d\eta^2 + d\vec x^2 )$
where $a$ is the expansion scale factor and $\eta$ is the conformal time. For
the purposes of CMB and mass power spectrum prediction, we   model the
$Q$-component of the fluid as a  scalar field, $Q$, with self-interactions
determined by a potential $V(Q)$. In an ideal adiabatic fluid with $w <0 $,  a
concern would be that the sound speed is imaginary, $c_s^2=w<0$, and small
wavelength perturbations are hydrodynamically unstable.\cite{Fabris}  However,
a scalar field is not an ideal fluid in this sense. The sound speed is a
function of wavelength, increasing from negative values for long wavelength
perturbations and approaching $c_s^2=1$ at small wavelengths (smaller than the
horizon size).\cite{Grishchuk} Consequently, for the cases considered here,
small wavelength modes remain stable. In general, our treatment relies only 
on properties of $w$ and $c_s^2$, and so it applies both to cosmic scalar
fields and to any other forms of matter-energy  ({\it e.g.}, non-scalar fields
or topological defects) with  $w$ and $c_s^2$ which can also be obtained by a
scalar field and some potential. Otherwise, we make no special assumptions
about the microscopic composition of the $Q$-component.

We implicitly assume that any couplings to other fields are negligibly small,
so that the scalar field interacts with other matter only gravitationally.
The  average  energy density is $\rho_Q = {1 \over 2 a^2} Q'^2 + V$ and  the
average pressure is $p_Q = {1 \over 2 a^2} Q'^2 - V$, where the prime
represents $\partial/\partial\eta$. As $Q$ evolves down its potential, the
ratio of kinetic (${1 \over 2 a^2} Q'^2$) to potential ($V$) energy can
change; this would lead to a time-varying equation-of-state, $w\equiv
p_Q/\rho_Q$. In this paper, we consider constant and time-varying $w$ models
for which  $w \equiv p_Q/\rho_Q \in [-1,0]$, since this range includes models
which best fit current observations.

We have modified numerical Boltzmann codes for computing CMB anisotropy and
the mass power spectrum. In one approach, we use an  {\it explicitly-defined
equation-of-state} $(w(\eta))$, which spans all models but requires that the
equation-of-state as a function of time be known.   All properties of the
model are determined by the proscribed  $w(\eta)\equiv w[a(\eta)]$ and 
\begin{equation}
\rho_Q = \Omega_Q \rho_o
\exp\Big[3 \log{a_o \over a} + 3\int_{a}^{a_o} \, {da \over a}  w(a)\, \Big],
\end{equation}
where $\rho_o$ and $a_o$ are respectively the critical energy density  and
scale factor today.  An effective scalar potential which produces this 
$\rho_Q$ is computed as a byproduct. A second approach uses an {\it
implicitly-defined equation-of-state} in which  we specify the potential
$V(Q)$ directly, which is useful  for exploring specific models  motivated by
particle physics.

A series of codes have been developed for synchronous  and for conformal
Newtonian gauge by modifying standard
algorithms.\cite{Crittenden,MaBertschinger,Seljak} In the synchronous
gauge, the line element is given by\cite{CDS}
\begin{equation}
ds^2 = a^2(\eta)[ -d\eta^2 + (\gamma_{i j} + h_{i j})dx^i dx^j]
\end{equation}
where $\gamma_{ij}$ is the unperturbed spatial metric, and $h_{ij}$ is the
metric perturbation. The scalar field fluctuation $\delta Q$ obeys the
equation
\begin{equation}
\delta Q'' + 2 {a' \over a}\delta Q' - \nabla^2\delta Q
+ a^2V_{,QQ}\delta Q = -{1 \over 2} h'Q' \, .
\label{synfluctsevol}
\end{equation}
Here, $h$ is the trace of the spatial metric perturbation, as described by Ma
and Bertschinger.\cite{MaBertschinger} 

We have found the observable fluctuation spectrum to be insensitive to a broad
range of initial conditions, including the case in which the amplitudes of
$\delta Q,\, \delta Q'$ were set by inflation. All of the examples shown in
this paper have $\delta Q=\delta Q'=0$ initially as measured in synchronous
gauge.  With this choice of initial conditions, the imprint of $\delta Q$  on
the CMB anisotropy would be negligible if we did not include  the response of
$Q$ to the matter density perturbations through the $h'Q'$ source term in 
Eq.~(\ref{synfluctsevol}). If $Q'=0$, as occurs for $\Lambda$CDM,  the scalar
field remains unperturbed. In the implicit approach where $V$ is given, the
energy density, pressure, and momentum perturbations are
\begin{eqnarray}
\delta\rho_Q &=& {1 \over a^2}Q'\delta Q' + V_{,Q}\delta Q, \qquad
\delta p_Q = {1 \over a^2}Q'\delta Q' - V_{,Q}\delta Q, \cr\cr
&&(\rho_Q + p_Q)(v_Q)_i = -{1 \over a^2} Q' (\delta Q)_{,i} \, .
\label{synflucts}
\end{eqnarray}
These quantities must be included in the evolution of $h$, the total
fluctuation density contrast $\delta$, and the total velocity perturbation
$v$. For cases where $w(\eta)$ is provided explicitly, the same relations may
be applied where an effective $V$ is  computed by the code from
$w(\eta)$.\cite{CDS}

Although we have examined  many models,\cite{CDS} we  restrict ourselves here
to  a representative spectrum. Figure 1 illustrates CMB studies showing
COBE-normalized\cite{Bunn} CMB anisotropy spectra  assuming adiabatic initial
conditions in the matter with tilt $n=1$, Hubble constant $H_0=100
h$~km~s$^{-1}$~Mpc$^{-1}$ with  $h=0.65$,  and $\Omega_b h^2 =0.02$, where
$\Omega_b$ is the ratio of the baryon density to the critical density.   
Because the models are spatially flat, the angular size of the  sound horizon
at last  scattering is nearly the same for all our models and, consequently, 
the acoustic oscillation peaks are at nearly the same $\ell$'s as in standard
CDM (SCDM).\cite{CMBReview,Hu}   However, the shape of the plateau  for low
$\ell$ and the shapes  of the acoustic peaks at high $\ell$ are distorted by
three effects: a  combination of early and late integrated Sachs-Wolfe
effects\cite{Hu}  due to having a fluid component with $w \ne 0$ and the direct
contribution of $\delta \rho_Q$     on the CMB.  Figures~1(a)-(b) show how the
spectra vary with $\Omega_Q$, the ratio of the energy density in $Q$ to the
critical density, for fixed $w$. The behavior is different for different
equations-of-state. For $w=-1/6$ the first acoustic peak rises and then 
steadily decreases as $\Omega_Q$ increases; for $w=-2/3$, the acoustic peak
rises uniformly as $\Omega_Q$ increases. Figure~1(c) compares predictions
assuming a smooth, non-fluctuating component (a case we argued is unphysical)
with the full spectrum including fluctuations in $Q$, illustrating the
substantial difference at large angular scales.  Figure 1(d) shows spectra for
fixed $\Omega_Q=0.6$ and varying $w$.  When fluctuations in $Q$ are not
included, the curves for $-1<w<-1/2$ are very nearly
degenerate,\cite{Steinhardt} but the degeneracy is  substantially broken in
Figure 1(d)  by the contribution of $Q$ fluctuations at  large angular scales 
at a level much greater than  cosmic variance.   Figure~1(e) illustrates 
results for a time-varying $w(\eta)$ obtained using
exponential\cite{RatraPeebles} and cosine\cite{CobleEtAl} scalar potentials. 
The exponential potentials are of the form $V(Q)=m^4 \, {\rm exp}[-\beta Q]$
where $Q=Q'=0$ initially;  for the examples shown, $(m,\, \beta)=(0.00358 \,
{\rm eV},\, 11.66 \, m_p^{-1})$ for the case with $w(\eta_0)=-1/6$ and $(m,\,
\beta)=(0.00243\, {\rm eV},\, 8.0 \, m_p^{-1})$  for $w(\eta_0)=-2/3$ as
expressed in  Planck mass ($m_p$) units. For the cosine potential, $V(Q)=m^4
(1+{\rm cos} [Q/f])$, where $(m,\, f)=(0.00465\, {\rm eV}, \, 0.1544 m_p)$ with
$Q=1.6 f$ and $Q'=0$  initially.\cite{CobleEtAl} Two of the curves are
especially interesting because they compare the CMB anisotropy for two models
with the same value of $w$ today but with different values of $w$ in the past.
One model has constant $w=-1/6$ whereas the other model, based on a scalar
field rolling down an exponential potential, has  $w$ evolving from -1 to 
-1/6. The two models produce easily distinguished  CMB power spectra even
though the equation-of-state is the same today due to the difference in past
evolutionary history and the direct contributions of $\delta Q$. Hence, the
evolution of $w$ can also be determined from CMB measurements. Also shown are
examples of other exponential and cosine potentials compared with SCDM which
illustrate that the CMB anisotropy is sensitive to  $V(Q)$. Figure~1(f)
illustrates an example  of CMB polarization, which has similar  variations with
parameters as the temperature anisotropy.

Figure~2 illustrates the mass power spectrum predictions. The key feature to
note in Figure 2(a) is that the power  spectrum is sensitive to all parameters:
the value of $w$; the time-dependence of $w$; the effective potential $V(Q)$
and initial conditions; and the value of $\Omega_Q$.  Hence, combined with the
CMB anisotropy, the power spectrum provides a powerful test of QCDM and its
parameters. An important  effect  is the suppression of the mass power spectrum
and $\sigma_8$ (the rms mass fluctuation at 8$h^{-1}$~Mpc) compared  to SCDM,
which makes for a better fit to current  observations.\cite{Ostriker} The grey
swath in Figure 2(b) represents the constraint from x-ray cluster abundance for
$\Lambda$CDM;\cite{Viana} in general, the constraint is weakly model-dependent.

While QCDM and $\Lambda$CDM both compare well to  current observations of CMB
and of large-scale structure today,  QCDM has  advantages in fitting
constraints from high red shift supernovae, gravitational lensing,  and
structure  formation at large red shift ($z\approx 5$). Constraints based on
classical cosmological tests on $\Lambda$ ($w_{\Lambda}=-1$), such as
supernovae and lensing, are significantly relaxed for QCDM with $w \approx
-1/2$ or greater.\cite{PJS} Another property of QCDM (or $\Lambda$CDM)  is
that structure growth and evolution ceases when the Q-component (or $\Lambda$)
begins to dominate over the matter density.  Comparing QCDM and $\Lambda$CDM 
models with $\Omega_Q= \Omega_{\Lambda}$, this cessation of growth occurs
earlier in QCDM.   For larger values of $w$, the  growth ceases earlier.
Hence,  more large scale structure and quasar formation at large red shift are
predicted,  in better accord with deep red shift images.

In conclusion, we find that  the ``quintessence" hypothesis fits all current
observations and results in an imprint on the CMB anisotropy and mass power
spectrum that should be detectable in near-future experiments. Its  discovery
could indicate the existence of new, fundamental fields with profound
implications for particle physics, as well as cosmology. A number of follow-up
studies  are underway including: a quantitative analysis to determine how well
future CMB experiments can resolve a  $Q$-component and how well one can
simultaneously resolve other cosmic parameters; numerical simulations of
galaxy formation and evolution in the presence of a $Q$-component; and  search
for quintessential candidates in models of fundamental physics.

We thank P.J.E. Peebles, J.P. Ostriker  and M. White for useful comments. This
research was supported by the Department of Energy at the University of
Pennsylvania, DE-FG02-95ER40893.

%%%%%%%%%%%%%%%%%%%%%%%%%%%%%%%%%%%%%%%%%%%%%%%%%%%%%%%%%%%%%%%%%%%%%%%%%%
%%%%%%%%%%%%%%%%%%%%%%%%%%%%%%%%%%%%%%%%%%%%%%%%%%%%%%%%%%%%%%%%%%%%%%%%%%

%%%%%%%%%%%%%%%%%%%%%%%%%%%%%%%%%%%%%%%%%%%%%%%%%%%%%%%%%%%%%%%%%%%%%%%%%%

\begin{figure}
\epsfxsize=3.3 in \epsfbox{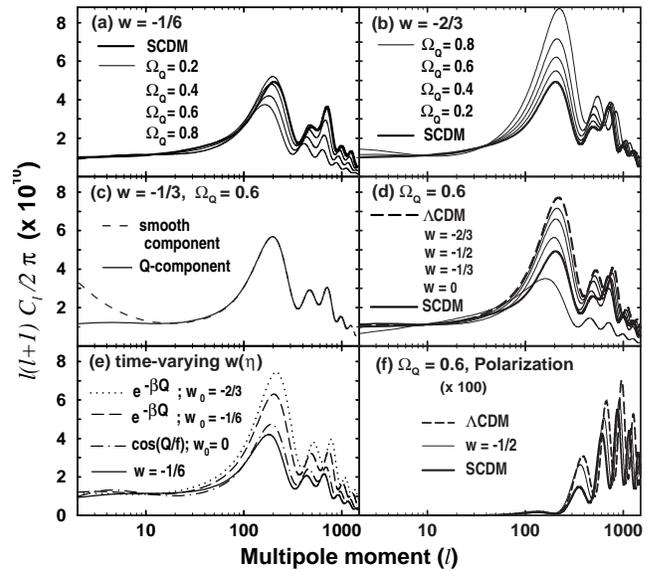}
\caption{CMB power spectrum, $\ell (\ell+1) C_{\ell}/2 \pi$ vs. $\ell$ where
$C_{\ell}$ is the multipole moment,  illustrating that the spectrum changes
significantly as a function  of  $w$ and $\Omega_Q$. Thick solid lines represent
SCDM; thick dashed lines correspond to   $\Lambda$CDM. For thin solid lines,
legends list the parameters  in sequence according to the height of the curve
beginning from the topmost. The variation with $\Omega_Q$ is shown in (a) and
(b);  (c) compares the predictions for a $Q$-component with fluctuations
($\delta Q \ne 0$) to  a smooth component  ($\delta Q$=0).  The 
differences due to $\delta Q$ at large angular scales  change 
COBE-normalization which  is responsible for
the substantial  variation of the acoustic  peaks with
$w$, as   shown in (d).  Panel (e) shows the results for time-varying $w(\eta)$ 
as determined by  specifying a potential $V(Q)$ and initial conditions,  where
$V(Q)$ is an exponential or cosine functional of $Q$ in the examples shown. 
Included also is the prediction for a constant $w=-1/6$, which  differs from the
results where $w(\eta)$ is time-varying  and $w(\eta_0)=-1/6$. In  (f) is shown 
the polarization (amplified 100-fold) for $w=-1/2$ compared to SCDM and
$\Lambda$CDM. }
\end{figure}

\begin{figure}
\epsfxsize=3.3 in \epsfbox{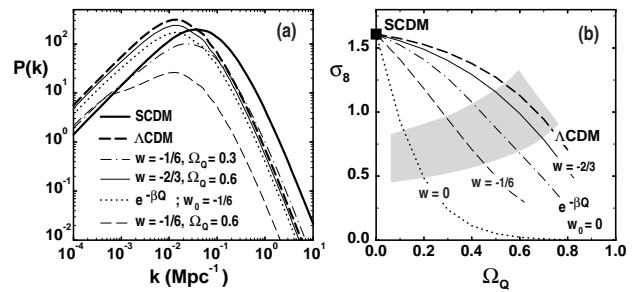}
\caption{ (a) Variation of  mass power spectrum  for some representative QCDM
examples. (b) The variation of $\sigma_8$ with $\Omega_Q$.   For $\Lambda$CDM,
$\Omega_Q$ is $\Omega_{\Lambda}$. The suppression of   $\sigma_8$ in  QCDM  
compared to  SCDM makes for a better fit with current observations.  The grey
swath illustrates constraints from x-ray cluster abundance.}
\end{figure}


\begin{references}

\bibitem{Ostriker}
See, for example,
J. P. Ostriker and P.J. Steinhardt, {\it Nature} {\bf 377}, 600 (1995),
and references therein.
 
\bibitem{Spergel} 
David Spergel  and Ue-Li Pen,
     Astrophys. J. 491, 67 (1997).

\bibitem{Steinhardt}
P. J. Steinhardt, in {\it Critical Dialogues in Cosmology},
ed. N. Turok, (World Scientific, 1997).

\bibitem{TurnerWhite}
Michael S. Turner and Martin White, {\it Phys. Rev. D}{\bf 56}, R4439 (1997).
 
\bibitem{CobleEtAl}
K. Coble, S. Dodelson, and J. A. Frieman,
{\it Phys. Rev. D}{\bf 55}, 1851 (1997).

\bibitem{Hill}
J. Frieman, C. Hill, A. Stebbins, and I. Waga, {\it Phys.
Rev. Lett.} {\bf 75}, 2077 (1995)

\bibitem{BestFitModelsPaper}
L. Wang, R. R. Caldwell,  J.P. Ostriker, and P.J. Steinhardt, in preparation.

\bibitem{SilveraWaga}
V. Silveira and I. Waga, {\it Phys. Rev. D}{\bf 56}, 4625 (1997).

\bibitem{Wetterich}
C. Wetterich, {\it Astron. Astroph.} {\bf 301}, 321 (1995).

\bibitem{RatraPeebles}
B. Ratra and P. J. E. Peebles, {\it Phys. Rev. D}{\bf 37}, 3406 (1988).

\bibitem{FerreiraJoyce}
Pedro G. Ferreira and Michael Joyce,
     Phys. Rev. Lett. 79, 4740 (1997).

\bibitem{Chiba}
Takeshi Chiba, Naoshi Sugiyama, and Takashi Nakamura,
     Mon. Not. R. Astron. Soc. 289, 5 (1997).

\bibitem{Fabris}
See, for example,
J. C. Fabris and J. Martin, {\it Phys. Rev. D}{\bf 55}, 5205 (1997).

\bibitem{Grishchuk}
L. Grishchuk, {\it Phys. Rev. D}{\bf 50}, 7154 (1994).
   
\bibitem{Crittenden}
R. Crittenden, J.R. Bond,  R.L.  Davis., G.  Efstathiou, and P.J. Steinhardt,
  {\it Phys.\ Rev.\ Lett.}{\bf 71}, 324 (1993).

\bibitem{MaBertschinger}
C.-P. Ma and E. Bertschinger, {\it Ap. J.} {\bf 455}, 7 (1995).

\bibitem{Seljak}
U. Seljak and M. Zaldarriaga, astro-ph/9603033.

\bibitem{CDS}
R.R. Caldwell, R. Dave, P.J. Steinhardt, in preparation.

\bibitem{Bunn} 
E.F. Bunn and M. White, {\it Ap. J.}{\bf 480},  6 (1997). 
 
\bibitem{CMBReview}
For reviews of CMB anisotropy power spectrum physics and properties,
see  P. J. Steinhardt, in {\it Proceedings of the Snowmass Workshop on the
Future of Particle Astrophysics and Cosmology}, ed. by E.W. Kolb and
R. Peccei, (World Scientific, 1995);
{\it IJMPA}  {\bf A}10, 1091-1124 (1995).

\bibitem{Hu} W. Hu, in {\it The Universe at High-z, Large Scale Structure and the 
Cosmic Microwave Background}, eds. E. Martinez-Gonzalez and J.L Sanz (Springer
Verlag); Wayne Hu, Naoshi Sugiyama, and Joseph Silk, {\it Nature} {\bf 386},
37 (1997).

\bibitem{Viana}
P.T.P. Viana and A.R. Liddle, 
{\it Mon. Not. Roy. Astron. Soc.} {\bf 281},  323  (1996).

\bibitem{PJS}  
P.J. Steinhardt, {\it Nature} {\bf 382}, 768  (1996).

\end{references}
\end{document}